\documentstyle[prb,aps,epsf,twocolumn]{revtex}
\newcommand{\eq}{\begin{equation}}
\newcommand{\ee}{\end{equation}}

\newcommand{\vx}{{\vec{x}}}

\def\lam{\lambda}

\def\half{{1\over2}}

\def\rhob{{\bar \rho}}

\def\eqa{\begin{eqnarray}}
\def\eea{\end{eqnarray}}
\def\prl{{Phys. Rev. Lett.}}
\def\prb{{Phys. Rev. {\bf B}}}

\begin{document}
\draft
\flushbottom
\twocolumn[
\hsize\textwidth\columnwidth\hsize\csname @twocolumnfalse\endcsname
\title{Temperature Dependence of the Spin Polarization in the Fractional
Quantum Hall Effects  }
\author{  Ganpathy Murthy}
\address{
Department of Physics and Astronomy, University of Kentucky, Lexington, KY 40506}
\date{\today}
\maketitle
\tightenlines
\widetext
\advance\leftskip by 57pt
\advance\rightskip by 57pt

\begin{abstract}
Using a Hamiltonian formulation of Composite Fermions that I recently
developed with R. Shankar, I compute the dependence of the spin
polarization on the temperature for the translationally invariant
fractional quantum Hall states at $\nu=1/3$ and $\nu=2/5$. I compare my
results to experiments at $\nu=1/3$, and find reasonably good agreement.

\end{abstract}
\vskip 1cm

]
\narrowtext
\tightenlines
The fractional quantum Hall (FQH) effect\cite{fqhe-ex} has introduced
us to new, highly correlated, incompressible states\cite{laugh} of
electrons in high magnetic fields.  A unified understanding of all
fractions $\nu=p/(2sp+1)$ was achieved by the Composite Fermion
picture of Jain\cite{jain-cf}, in which the electrons are dressed by $2s$
units of statistical flux to form Composite Fermions
(CFs). At a mean field level, the CFs  see a reduced field
$B^*=B/(2sp+1)$, in which they fill $p$ CF-Landau levels (CF-LLs), and
exhibit the integer quantum Hall effect. 

Due to the small $g$ factor of electrons in GaAs, spins may not be
fully polarized in FQH states\cite{hal-spin,singlet25}. Transitions
between singlet, partially polarized, and fully polarized states
(based on gap measurements) have been observed for a number of
fillings\cite{clark,initialex,buckthought,duetal}, which can be
understood in terms of CF's with a spin\cite{jain-cf,duetal,wu}. The
transitions happen when an unoccupied CF-LL of one spin crosses the
occupied CF-LL of the opposite spin. 

While these low temperature measurements are in satisfactory agreement
with the ground states predicted in the CF picture\cite{wu}, in order
to understand the temperature dependence of the polarization $P(T)$
one has to consider all excited states as well. Detailed measurements
of $P(T)$ for the $\nu=1/3$ state have recently appeared in the
literature\cite{barrett,melinte}. It is well-known that the $\nu=1/3$
state is spontaneously polarized at $T=0$, even when the Zeeman
coupling $E_Z=g\mu B_{tot}$ is zero. In this it is analogous to the
$\nu=1$ state\cite{shivaji-skyrmion}, which has been extensively
studied theoretically\cite{largen,many-body,haussmann} and
experimentally\cite{skyrmion-ex}. There are, however, significant
differences between the two cases at finite $T$. The $P(T)$ curve for
$\nu=1/3$ has a different shape, and has empirically been fitted to a
{\it noninteracting} form $P(T)=\tanh\big(\Delta/4k_BT\big)$ where
$\Delta$ is found to approximately twice $E_Z$.

In a recent paper, MacDonald and Palacios\cite{macd-pala} identified a
key qualitative feature that makes $\nu=1/3$ very different from
$\nu=1$. In the $\nu=1$ case the particle-hole excitations are very
high in energy compared to $E_Z$, and are frozen out at all low
temperatures of interest. Consequently, the $T$ dependence of $P$
comes mainly from spin wave excitations and their
interactions. This is the reason why long-wavelength effective
theories such as the continuum quantum ferromagnet\cite{largen} 
approach are successful. However, for $\nu=1/3$, particle-hole
excitations are on the same scale of energy as $E_Z$, and cannot be
ignored at any $T$. MacDonald and Palacios use a simplified model to
illustrate this feature\cite{macd-pala}, but the model is not
sufficiently detailed to enable a calculation of $P(T)$ for a
realistic sample. Also, the model cannot be readily  extended to
non-Laughlin fractions.

The goal of this paper is to describe a general analytical method for
approximately computing $P(T)$ for an {\it arbitrary} principal
fraction for realistic samples.  In the last few years, R. Shankar and
I have developed a Hamiltonian formalism\cite{us1} which allows us to
carry out approximate calculations for any physical quantity in the
fractional Hall regime. Our central result is a formula for the
LLL-projected electronic charge density at small $q$:
\eq
\rho_e(q)={\sum_j e^{-iqx_j} \over 2ps+1} -{il_{}^{2} }  (\sum_j (q \times
\Pi_j)e^{-iqx_j}
)\label{rhobar}
\ee
where $\vx_j$ is a CF coordinate, $l=1/\sqrt{eB}$ is the magnetic
length, and ${\vec\Pi}_j={\vec P}_j+e{\vec A}^*(r_j)$ is the velocity
operator of the CFs.  The low-energy Hamiltonian is $ H=\half \int
{d^2 q\over(2\pi)^2} v(q)
\rhob(-q) \rhob(q) $ where $v(q)$ is the electron-electron
interaction. To include the effects of finite sample thickness, and to
stay within the limitations of our small-$q$ approach, we work with a
modified Coulomb interaction of the form $v(q)=e^{-\lam q}2\pi e^2/q$,
where the length $\lam$ is connected to the thickness.  A notable
feature of the formalism is that energy dispersions arise entirely
from interaction effects\cite{us1}. The Hartree-Fock (HF)
approximation has been applied to the above Hamiltonian, and
reasonable success has been obtained in computing
gaps\cite{single-part} (and scaling relations between
gaps\cite{scaling}) for the principal fractions, and magnetoexciton
dispersions\cite{me-us}. Most recently Shankar has computed  $P(T)$ for
the compressible half-filled LL\cite{shankar12}. The reason HF works
so well is that our Hamiltonian is expressed directly in terms of CF
coordinates. The CF's have the right (fractional) charge and dipole
moment in our formalism, and corrections to HF are expected to be
small for most physical quantities.

We will compute $P(T)$ for $1/3$ and $2/5$ by simply carrying out the
Hartree-Fock (HF) approximation for CFs at finite $T$. We will use a
CF-LL cutoff to ensure that the correct number of electronic states
occur in the Hilbert space. This implies that one must keep 3 CF-LLs
for $\nu=1/3$ and 5 CF-LLs for $\nu=2/5$. While this restriction is
relatively unimportant at very low $T$, it becomes increasingly
important as the temperature increases, and occupations of the excited
CF-LLs become significant.

Let us proceed to the results. We first consider the 10W sample of
Khandelwal {\it et al}\cite{barrett}. The sample parameters are
$B_{\perp}=9.61 T$, $B_{tot}=12T$, and each quantum well has a
thickness of 260$\AA$. This implies that the Coulomb energy scale is
$E_C=e^2/\varepsilon l_0\approx160 K$ and the Zeeman energy is $E_Z=0.0175
E_C$. The nominal thickness of the sample is $\lambda\approx
3l_0$. 

\begin{figure}
\narrowtext
\epsfxsize=2.4in\epsfysize=2.4in
\hskip 0.3in\epsfbox{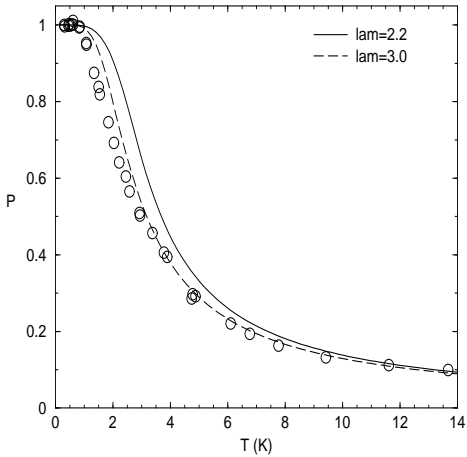}
\vskip 0.15in
\caption{Polarization versus $T$ for $\nu=1/3$. The circles are the 
data from the 10W sample of Khandelwal {\it et al}\cite{barrett}. The
solid line is the prediction from our theory for $\lambda=2.2l_0$, while
the long dashed line is for $\lambda=3.0l_0$.
\label{fig1}}
\end{figure}

Figure 1 shows the HF prediction from our theory for $\lambda=3l_0$
compared to the experimental data (dashed line). The agreement is very
gratifying. However, it must be regarded as fortuitous, since the
simple model potential that we have assumed is unlikely to reproduce
the complicated effects of density distributions in the actual sample
with finite thickness. In other words, $\lambda$ is related to the
sample thickness only in a complicated and indirect way. Another way
to determine the value of the effective $\lambda$ is to go to the
calculation of Shankar\cite{shankar12} for $\nu=1/2$, in which he
found a reasonable fit to $P(T)$ assuming $\lambda\approx1.75l_0$ for
the {\it same} 10W sample. Accounting for the fact that the magnetic
length changes when one changes filling at constant density, we
estimate $\lambda\approx2.2l_0$ for $\nu=1/3$. Figure 1 also shows the
prediction for this value (solid line). It can be seen that the
predicted curve lies above the data over a range of intermediate
$T$. This is only to be expected since the simple HF does not include
the effects of spin wave excitations, or of the modification of
single-particle energies due to the interaction of CFs with spin
waves.  The agreement between theory and experiment shown in Figure 1
is reasonably good, even for $\lambda=2.2$. It can also be seen that
changes in $\lambda$ do not make huge changes in $P(T)$.

\begin{figure}
\narrowtext
\epsfxsize=2.4in\epsfysize=2.4in
\hskip 0.3in\epsfbox{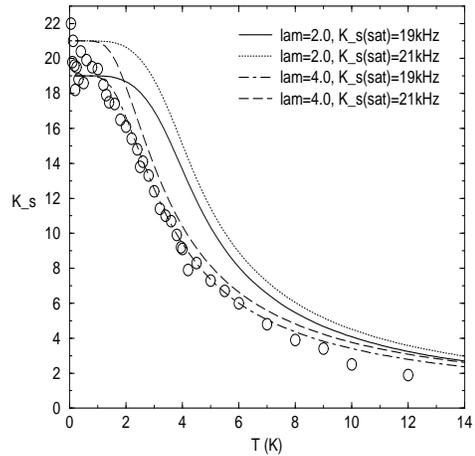}
\vskip 0.15in
\caption{Knight shift versus $T$ for $\nu=1/3$. The circles are the 
data from the M242 sample of Melinte {\it et al}\cite{melinte}. The
 lines are the predictions from our theory for $\lambda=4l_0$ and
 $2.0l_0$, assuming two different values for the Knight shift that
 corresponds to $P=1$.
\label{fig2}}
\end{figure}

Figure 2 shows the same type of comparison for the data of Melinte
{\it et al}\cite{melinte}for their M242 sample.  Here the sample
parameters are $B_{tot}=B_{\perp}=17T$, and the nominal well thickness
is 250$\AA$. This implies that $E_C\approx210K$, $E_Z=0.0186
E_C\approx4K$, and a nominal value for $\lambda=4.0l_0$. Once again
this value for $\lambda$ is very likely an overestimate, so I have
also calculated the prediction for $\lambda=2.0l_0$. There is a lot of
scatter in the data at low $T$, due to the very long times needed to
measure the Knight shift, and the error bars are also large at low
$T$\cite{melinte}. This gives us some latitude in defining what we
mean by the Knight shift corresponding to $P=1$. In any reasonable
theory one expects to find that $P=1$ for $T\ll E_Z$, and expects to
see this saturated value of $P$ up to about $T=0.5E_Z$ or so.

Based on these considerations I have used two values of $K_{s,P=1}$,
$21kHz$ and $19kHz$ both of which lie within the error bars of the low
$T$ data\cite{melinte}. One possibility that can explain this spread
is that spin-reversed quasiparticles are present in the ground state
due to disorder, which can bring down the ``saturated'' value of the
Knight shift\cite{jain-pvt}. The $21kHz$ value was used by Melinte
{\it et al} in a phenomenological $\tanh(\Delta/k_BT)$ fit to obtain
$\Delta=1.7E_Z$. I believe that the fit for $K_{s,P=1}=19kHz$ is
closer to the truth, since then the experimental saturation region is
about $0.5E_Z$. The agreement between theory and experiment for this
value of $K_{s,P=1}$ are much better than for
$K_{s,P=1}=21kHz$. Overall the agreement is somewhat worse than for
the Khandelwal {\it et al} data\cite{barrett}, but still adequate,
considering the simple nature of the approximation.

Why is HF so good in this case while it was so poor\cite{many-body} for
$\nu=1$? To answer this question let us turn to the
spin wave dispersions. These can be computed in the manner described
in my magnetoexciton calculation\cite{me-us}, and are shown in Figure
3 for $\lambda=2.2l_0$ and $3.0l_0$ for $E_Z=0.0175 E_C$ and $T=0$. The $q\to
0$ limit is required to be $E_Z$ by Larmor's theorem, while the
$q\to\infty$ limit is the spin-reversed particle-hole gap
$\Delta_{SR}$. Figure 3 explicitly illustrates the
feature\cite{macd-pala} that the spin-flip particle-hole excitations
are at the same energy scale as $E_Z$. This in turn points to the need
for a more sophisticated theory of these ferromagnets which includes
particle-hole excitations, spin waves, and their interactions with
themselves and each other. Such a complete theory does not yet exist.
\begin{figure}
\narrowtext
\epsfxsize=2.4in\epsfysize=2.4in
\hskip 0.3in\epsfbox{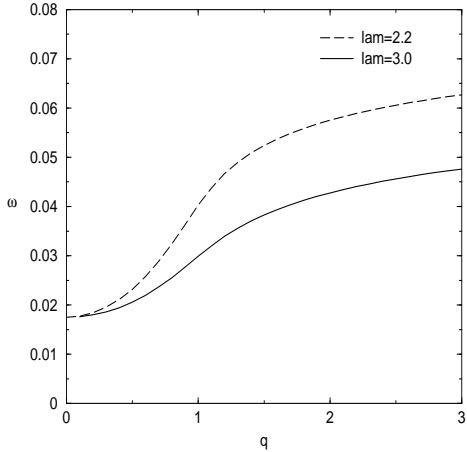}
\vskip 0.15in
\caption{Spin wave energy dispersions in units of the Coulomb energy $E_C$  
for $E_Z=0.0175$, and $\lambda=2.2l_0$ and $3.0l_0$. As can be seen,
the scale of the spin-reversed gap is the same as $E_Z$.
\label{fig3}}
\end{figure}

Figure 4 shows the evolution of $\Delta_{SR}$ with $T$. Recall that
all the energy splittings in our theory come from interactions, and as
the occupations of the states change with $T$ so do their energies. It
is clear that as $T$ becomes large $\Delta_{SR}$ tends rapidly towards
$E_Z$. This implies that the finite $T$ spin wave dispersion becomes
increasingly flat at $T$ increases. Recall that a {\it noninteracting}
theory would have a completely flat dispersion, that is,
$\omega(q)=E_Z$. 
\begin{figure}
\narrowtext
\epsfxsize=2.4in\epsfysize=2.4in
\hskip 0.3in\epsfbox{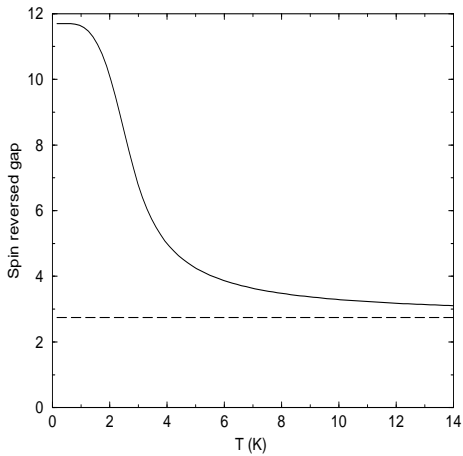}
\vskip 0.15in
\caption{Spin-reversed gap as a function of $T$ (both in units of $E_C$). 
The dashed line is $E_Z$.  For $T\ge 5K$ the theory is essentially
noninteracting.
\label{fig4}}
\end{figure}
Therefore, as the temperature increases, our theory
becomes more weakly interacting, and our HF becomes more
accurate. This trend can be expected to continue until a temperature
scale when CFs cease to exist. There are no obvious signs of such a
scale in the data.

Our theory is very general, and can be applied to any fractional Hall
state. To ilustrate this Figure 5 shows the $P(T)$ curves for
$\nu=2/5$ for $\lambda=1.5l_0$ for a range of Zeeman couplings. Note
the nonmonotonicity of the curves that start from the singlet ground
state at $T=0$. There is a transition to the fully polarized state
around $E_Z=0.01E_C$. Note also that I have allowed only
translationally invariant HF states, which ignores possible partially
polarized states that I have recently proposed\cite{ppdw} to explain
intriguing observations by Kukushkin {\it et al}\cite{kukush} of a
state with half the maximal polarization for $\nu=2/5$, which is not
allowed as a translationally invariant CF state. I plan to explore the
temperature dependence of the polarization, and other properties of
this inhomogeneous state more thoroughly in a future publication.

\begin{figure}
\narrowtext
\epsfxsize=2.4in\epsfysize=2.4in
\hskip 0.3in\epsfbox{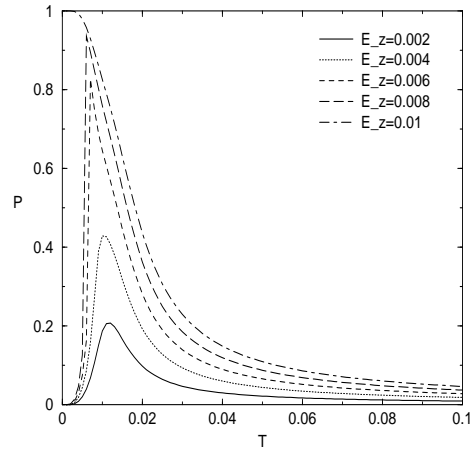}
\vskip 0.15in
\caption{Polarization versus $T$ for $\nu=2/5$ and $\lambda=1.5l_0$, 
for a range of values of $E_Z$. $E_Z$ and $T$ are both plotted in
units of $E_C$.
\label{fig5}}
\end{figure}

Finally, let us compare our results to the only other method that can
compute $P(T)$ for arbitrary fractions, which is exact diagonalization
(keeping all the excited states) and subsequent calculation of
thermodynamic quantities\cite{exact}. Due to computational
limitations, this method is restricted to fairly small systems. For
example, the largest system studied by Chakraborty and
Pietilainen\cite{exact} for $\nu=1/3$ has 5 electrons, and for
$\nu=2/5$ has 4 electrons.  For $\nu=1/3$ the exact diagonalization
result lies above our predictions (and the experiment) for $T>4K$. In
fact, at $T=0.09E_C\approx 14K$, the exact diagonalization prediction
seems to be almost a factor of two above our prediction, which
essentially coincides with the experiment (Figure 1). This discrepancy
might be the result of finite thickness or finite size
corrections. However, at low $T$ the exact diagonalization
result\cite{exact} follows the data more closely than our HF
approximation (in all the above comparisons I have used the $g=0.5$
line in Figure 2 of ref[27] and compared to the 10W sample of
Khandelwal {\it et al}\cite{barrett}. This sample has the closest
parameters to those used in ref[27]). For $\nu=2/5$, our results
reproduce the nonmonotonicity of $P(T)$ for those values of $E_Z$
where the singlet state is the ground state, and the peaks in $P(T)$
occur at roughly the same $T$ in our results and the exact
diagonalization results\cite{exact}. However, the same overall pattern
holds for $\nu=2/5$, namely,  the results of Chakraborty and
Pietilainen\cite{exact} are below ours for low $T$, but are higher for
$T>0.02E_C$, where they once again see a $1/T$ tail with a large
coefficient. It would be interesting to explore the finite size
systematics to see if the large $T$ tail is suppressed for larger
sizes.

In summary, I have presented an approximate analytical method for
computing the temperature dependence of the polarization for an
arbitrary fractional quantum Hall state.  An important open problem is
the development of a formalism that can successfully compute
corrections to HF by including particle-hole excitations, spin waves,
and their interactions in a self-consistent way at finite $T$. While
the problem is not pressing for $\nu=1$, where large-$N$
treatments\cite{largen} give quite good agreement with experiment, it
is sorely needed for fractional Hall ferromagnets. It will be
interesting to see whether such corrections to HF can improve the
agreement between our predictions and experiment.

It is a pleasure to thank R. Shankar, J. K. Jain, S. E. Barrett, and
N. Freytag for helpful conversations, and the latter two for sharing
their raw data.

\end{document}